\documentstyle[11pt,fleqn,cite]{article}   
\catcode`\@=11
\@addtoreset{equation}{section}
\def\theequation{\arabic{section}.\arabic{equation}}
\def\appendix{\renewcommand{\thesection}{\Alph{section}}\setcounter{section}{0}
              \renewcommand{\theequation}
            {\mbox{\Alph{section}.\arabic{equation}}}\setcounter{equation}{0}}
\def\maketitle{\thispagestyle{empty}\setcounter{page}0\newpage
                \renewcommand{\thefootnote}{\arabic{footnote}}
                  \setcounter{footnote}0}
\renewcommand{\thanks}[1]{\renewcommand{\thefootnote}{\fnsymbol{footnote}}
               \footnote{#1}\renewcommand{\thefootnote}{\arabic{footnote}}}
\newcommand{\preprint}[1]{\hfill{\sl preprint - #1}\par\bigskip\par\rm}
\renewcommand{\title}[1]{\begin{center}\Large\bf #1\end{center}\rm\par\bigskip}
\renewcommand{\author}[1]{\begin{center}\Large #1\end{center}}
\newcommand{\address}[1]{\begin{center}\large #1\end{center}}

\def\dinfn{\smallskip Dipartimento di Fisica, Universit\`a di Trento\\ 
                           and Istituto Nazionale di Fisica Nucleare,\\
                                   Gruppo Collegato di Trento, Italia}

\def\Idinfn{\address{\dinfn}}
\def\dinbcn{\smallskip 
Consejo Superior de Investigaciones Cient\'{\i}ficas \\
IEEC, Edifici Nexus 201, 
                                Gran Capit\`a 2-4, 08034 Barcelona, Spain\\
                   and Departament ECM and IFAE, Facultat de F\'{\i}sica, \\
              Universitat de Barcelona, Diagonal 647, 08028 Barcelona, Spain}
\def\Idinbcn{\address{\dinbcn}}
\newcommand{\email}[1]{e-mail: \sl #1@science.unitn.it\rm}
\newcommand{\zmail}[1]{e-mail: \sl #1@zeta.ecm.ub.es\rm}
\newcommand{\fzmail}[1]{\thanks{\zmail{#1}}}
\newcommand{\femail}[1]{\thanks{\email{#1}}}
\newcommand{\pacs}[1]{\smallskip\noindent{\sl PACS numbers:
                       \hspace{0.3cm}#1}\par\bigskip\rm}
\def\babs{\hrule\par\begin{description}\item{Abstract: }\it} 
\def\eabs{\par\end{description}\hrule\par\medskip\rm}
\renewcommand{\date}[1]{\par\bigskip\par\sl\hfill #1\par\medskip\par\rm}
\newcommand{\ack}[1]{\par\section*{Acknowledgments} #1} 
\newcommand{\s}[1]{\section{#1}}
\renewcommand{\ss}[1]{\subsection{#1}}

\renewcommand{\vec}[1]{{\bf #1}}       %%%  vectors in bold
\newcommand{\ca}[1]{{\cal #1}}         %%%  calligraphic
\def\hs{\qquad}               %%%  horizontal space
\def\nn{\nonumber}            %%%  no number for eqnarray
\def\beq{\begin{eqnarray}}    %%%  begequation/eqnarray
\def\eeq{\end{eqnarray}}      %%%  endequation/eqnarray
\def\ap{\left.}               %%%  open bracket
\def\at{\left(}               %%%  open (
\def\aq{\left[}               %%%  open [
\def\ag{\left\{}              %%%  open {
\def\ct{\right)}              %%%  close )
\def\cq{\right]}              %%%  close ]
\def\cg{\right\}}             %%%  close }
\def\R{{\hbox{{\rm I}\kern-.2em\hbox{\rm R}}}}   %%% real numbers
\def\H{{\hbox{{\rm I}\kern-.2em\hbox{\rm H}}}}   %%% Hilbert space
\def\N{{\hbox{{\rm I}\kern-.2em\hbox{\rm N}}}}   %%% natural numbers
\def\C{{\ \hbox{{\rm I}\kern-.6em\hbox{\bf C}}}} %%% complex numbers
\def\Z{{\hbox{{\rm Z}\kern-.4em\hbox{\rm Z}}}}   %%% integers numbers
\def\ii{\infty}                                  %%% infinit
\newcommand{\fr}[2]{\mbox{$\frac{#1}{#2}$}}      %%% small fraction
\def\tr{\mbox{ tr}\ }                  %%% trace
\def\Tr{\mbox{ Tr}\ }                  %%% Trace
\def\Res{\mbox{ Res}\ }                %%% Residue
\def\res{\mbox{ res}\ }                %%% Residue
\renewcommand{\Re}{\mbox{ Re}\ }       %%% Real 
\def\lap{\Delta}                                   %%% Laplacian
\def\al{\alpha}
\def\be{\beta}
\def\ga{\gamma}
\def\de{\delta}

\def\ze{\zeta}

\def\la{\lambda}

\def\si{\sigma}
\def\om{\omega}
\def\ph{\varphi}

\def\Ga{\Gamma}

\begin{document}
%\tableofcontents       %%%%%%   index of section

\preprint{UTF 413}

\title{Applications in physics of the multiplicative anomaly formula \\
involving some basic differential operators}

\author{Emilio Elizalde\fzmail{eli}}
\Idinbcn
\author{Guido Cognola\femail{cognola}
and Sergio Zerbini\femail{zerbini} }
\Idinfn

\date{April 1998}

\babs
In the framework leading to the multiplicative anomaly formula
---which is here proven to be valid even
in cases of known spectrum but non-compact manifold (very important in
Physics)--- zeta-function regularisation techniques are shown to be
extremely efficient.
Dirac like operators and harmonic oscillators are investigated in
detail, in any number of space dimensions. They  yield a
non-zero anomaly which, on the other hand, can always be expressed
by means of a simple analytical formula.
These results are used in several physical examples,
where the determinant of a product of differential operators is not
equal to the product of the corresponding functional determinants.
The simplicity of the Hamiltonian operators chosen is aimed at showing
that such situation may be quite widespread in mathematical physics. 
However, the consequences of the existence of the determinant 
anomaly have often been overlooked.
 \eabs



\newpage

\s{Introduction}
\label{Form}

The importance of zeta-function regularisation for the definition
 of functional determinants \cite{ray71-7-145} is, without discussion,  a
powerful tool to deal with the ambiguities (ultraviolet divergences) present
 within the one-loop or external field approximation, in  relativistic quantum
field theory (see the seminal papers \cite{dowk76-13-3224,hawk77-55-133} and
for  recent reviews\cite{eliz94b,byts96-266-1}). Let us remember that 
 the Euclidean partition related to a 
quantum scalar field can be formally written as
\beq
Z=\at \det L_D \ct^{-1/2}
\:,\label{0}\eeq
with $L_D$ an elliptic differential operator. The latter quantity is 
ill defined and, with regard to this, we briefly recall
 how zeta-function regularisation works: 
it gives a precise
meaning, in the sense of analytic continuation,
to the determinant of a differential operator which, as the product of its
eigenvalues, is formally divergent.
When the ($D$-dimensional) manifold is smooth and compact,
 the spectrum is discrete and one has,
for $\Re s>D/l$,
$l$ being the order of the differential operator $L_D$,
the definition of the zeta-function
\beq
\ze(s|L_D)=\sum_i\la_i^{-s}
\:,\nn\eeq
where $\la_i$ are the eigenvalues of the elliptic operator $L_D$.
Making use of the relationship between the zeta-function and the
heat-kernel trace via the the Mellin transform, when $\Re s>D/l$, one can write
\beq
\ze(s|L_D)=\Tr L_D^{-s}
=\frac{1}{\Ga(s)}\int_0 ^\ii t^{s-1}
\:K(t|L_D)\:dt
\,,
\eeq
where $K(t|L_D)=\Tr\exp(-t L_D)$ is the trace of the heat operator.
The previous relations are valid also in the presence of
zero modes, with the replacement
$K(t|L_D)\longrightarrow K(t|L_D)-P_0$,
$P_0$ being the projector onto the set of zero modes.

A well known heat-kernel expansion argument leads to the 
meromorphic structure of $\zeta(s|L_D)$. 
It is found that the analytically continued zeta-function is regular
at $s=0$ and thus its derivative in that point is well defined.
As a consequence, the one-loop Euclidean partition function, regularised by
means of the zeta-function, reads \cite{hawk77-55-133}
\beq
\ln Z=-\frac{1}{2}\ln\det(\ell^2 L_D)
=\frac{1}{2}\ze'(0|\ell^2 L_D)
=\frac{1}{2}\ze'(0| L_D)-\frac{1}{2}\ze(0|L_D)\ln\ell^2
\nn\:,\eeq
where
$\ze'$ is the derivative with respect to $s$ and
$\ell$ is a renormalization scale.

However, in general things are not so simple. Sometimes it happens
that one has to deal with the product of two or more
differential operators and one is directly confronted 
with the validity (or not) of the multiplicative property:
\beq
\ln\det (AB) =\ln\det A+\ln\det B
\:,\label{ab}\eeq
which is of course known to hold for non singular matrices in finite
dimensional vector spaces.

A first elementary example is the following. Consider a
free vector-valued scalar field $\phi_i$  in $R^4$, with a broken
$O(N)$ symmetry ---owing to the mass terms $m_i^2$. The Euclidean action is
\beq
S=\int  dx^4 \phi_i \aq \at-\lap+m^2_i \ct \phi_i
\cq
\:,\label{ea}\eeq
and the related Euclidean operator
reads
\beq
L_{ij}=\at -\lap+m^2_i
\ct\de_{ij}
\:,\label{sd}\eeq
in which $\lap$ is the Laplace operator. Thus, one is actually
dealing with a matrix-valued elliptic differential operator. In
this case,  the Euclidean partition function is given by
\beq
\ln Z=-\frac{1}{2}\ln\det\left|\ell^2L_{ij}\right| 
=-\frac{1}{2}\ln\det\aq\ell^2(-\lap+m^2_1)...\ell^2(-\lap+m^2_N)\cq
\,.\label{02}
\eeq

Another, less trivial, example concerns the finite temperature effects
for a gas of free relativistic charged bosons. With regard to this
case, it is possible to show
that the logarithm of the grand canonical partition function, choosing 
as parametrization of the charged boson field the two real scalar fields 
$\phi_i$, can be expressed by \cite{bens91-44-2480}
\beq
\ln Z_{\be,\mu}=-\frac{1}{2}\ln\det\at\ell^4 L_+ L_-\ct
\:,\label{mu}\eeq
with
\beq
L_\pm=-\partial_\tau^2-\lap+m^2+\mu^2\pm 2\mu \sqrt{-\lap+m^2}
 \:,\label{ll}\eeq
$\tau$ being the imaginary time compactified with period $\be$, the inverse
of equilibrium temperature, and $\mu$ the chemical potential. Note 
however that, if one chooses as parametrization of the charged boson 
field the complex scalar fields 
$\phi$ and $\phi^*$, one has 
\beq
\ln Z_{\be,\mu}=-\frac{1}{2}\ln\det\at\ell^4 K_+ K_-\ct
\:,\label{mu2}\eeq
with
\beq
K_\pm=-\partial_\tau^2-\lap+m^2+\mu^2\pm 2\mu \partial_\tau  
 \:.\label{llk}\eeq

In a recent work, Stuart Dowker \cite{dowk98u-200} has shed doubt on 
the recipe used in computing the  partition functions, Eqs.~(\ref{02}) and 
(\ref{mu}). In our opinion, the above prescription is correct and we refer 
to Ref.~\cite{eliz98u-72}, where this issue has been discussed in detail.

As further examples, we simply recall that in the evaluation of the
one-loop Vilkovisky-DeWitt effective action
(see, for example, 
Refs.~\cite{vilk84-234-125}--\cite{bala92-46-4413}
for details)
and in some GUT-like models
(see, for example, \cite{cogn95u-366}), one has to deal with matrices
of higher order differential operators,
which give rise to products of the same,
when one has to compute  functional determinants.

Within the physical literature,
in all the examples we have recalled, and in many other that
 we do not mention here,
the way one usually proceeds is by
formally assuming the validity of the multiplicative
relation, indiscriminately.
Needless to say, this may be dangerous. One has to use always
some regularisation procedure and
it turns out, in fact, that the regularised determinants
do {\it not} satisfy in general the multiplicative property, Eq.~(\ref{ab}).
Even when one is dealing with commuting operators,
there exists the so-called  multiplicative
anomaly \cite{kass89-177-199,kont95b}.
In terms of $F(A,B)\equiv\det(AB)/(\det A\det B)$ \cite{kont95b},
it is simply defined as
\beq
a(A,B)=\ln F(A,B)=\ln \det (AB)-\ln \det (A)-\ln \det (B)
\:.\label{ma}\eeq

It should be noted by passing that the
non vanishing of the multiplicative anomaly implies that the relation
\beq
\ln \det A =\Tr \ln A
\label{tl}\eeq
does not hold, in general, for elliptic operators. The formal use
of the above operator identity is not justified, since the
multiplicative anomaly may be present. In fact, if one assumes that
$\Tr\ln A$ is a linear functional, in the simplest case of
$[A,B]=0$, one has
\beq
\Tr \ln AB=\Tr (\ln A+\ln B)=\Tr \ln A+\Tr \ln B
\label{tl1}
\,.\eeq
Thus, if Eq.~(\ref{tl}) holds, one arrives at  a contradiction with
Eq.~(\ref{ma}), as soon as the multiplicative anomaly is not
vanishing. With regard to this issue, taking  for granted 
Eq.~(\ref{tl}), one might start with the {\it 
definition} 
\beq
Z=\exp\left(-\frac{1}{2}\Tr \ln L_D\right)
\:,\label{01}\eeq
instead of Eq.~(\ref{0}) and then make use of some regularisation 
while preserving the linear property of the trace. In this case, in 
Ref. \cite{evan98u-184} the absence of the multiplicative 
anomaly has been claimed and doubt has been shed  on the use of 
zeta-function regularisation. In our opinion, this is not justified (see 
Ref. \cite{eliz98u-72a}).

The presence of the multiplicative anomaly in the commuting case
might turn out as a surprise. However,
in Ref. \cite{eliz97u-404} it has been shown that its existence is actually
unavoidable
in order to have independence of the partition function on the choice of
bosonic degrees of freedom (see, also \cite{eliz98u-72}). 
In some cases, the multiplicative anomaly gives a
contribution to the
effective action which can be readily absorbed into the renormalization
procedure,
but in some systems it may certainly  produce physical consequences,
which have to
be carefully analysed. A non trivial example has been discussed in
Ref.~\cite{eliz97u-404}, where after a renormalization of the charge
operator, it has been shown that, starting from the natural definition  
Eq.~(\ref{0}), the multiplicative anomaly gives rise
to a new contribution ---overlooked in  previous treatments--- to the high
temperature expansion of the free energy of the relativistic boson gas
in the symmetric unbroken phase.

We will show, and illustrate with the help of several examples, that this
 multiplicative anomaly appears already in very simple situations
(one-dimensional, first order differential operators differing in a constant
term). And that it can be most conveniently
 expressed by means of the non-commu\-ta\-ti\-ve residue
associated with a classical pseudo-differential operator, known as the
Wodzicki residue \cite{wodz87b}.
In fact, the purpose of the present paper is to continue the analysis
of the emergence of the multiplicative anomaly, that was started in
Refs.~\cite{eliz97u-394,eliz97u-404}
---where the commutative case was
discussed--- and to extend it to the more general non-commutative case,
working out explicitly the multiplicative
anomaly formula for a large class of elliptic operators ---and to investigate
some very basic examples with care. In particular, in some of these
examples we will deal with  non compact manifolds, where, strictly
speaking, the Wodzicki theory has not  yet been developed.
Nevertheless, we will see in such examples that the formula
for the multiplicative anomaly turns out to be valid too.

The content of the paper is as follows.
In Sect.~\ref{S:PDMA}  a perturbative derivation of the
multiplicative anomaly for a particular class of differential operators
in arbitrary dimensions is presented, while the general formula in the case
of lower dimensions $D\leq4$ is given in \ref{S:EEAPT}.
In Sect.~\ref{S:WR} we shall recall the definition of the Wodzicki residue,
together with
some related results, which will be used in Sect.~\ref{S:MAF}
in order to render explicit its relation with the multiplicative anomaly
formula. Then, we show in \ref{S:MALD} that such a general formula
can be notably simplified in lower dimensions,
confirming completely the result obtained by using perturbation theory.
In Sect.~\ref{S:HKFA}, we show how the multiplicative anomaly might be used in order to
compute the heat-kernel coefficients.
Finally, in Sect.~\ref{S:examples} we 
discuss some basic physical examples, formulated in
non-compact manifolds, but for which the spectrum of the Hamiltonian operator
is exactly known, and we  compute the anomaly explicitly.
The paper ends with some conclusions in Sect.~\ref{S:conclusion}.

\s{Perturbative derivation of the the multiplicative anomaly}
\label{S:PDMA}

In this section we will consider a particular case
involving  two self-adjoint elliptic invertible
operators $H$, $H_V=H+V$, and the related product
\beq
A=H(H+V)\:.
\eeq
By definition, the corresponding multiplicative anomaly reads
\beq
a(H,H_V)=\ln\det A-\ln\det H-\ln\det H_V\,
\eeq
where the functional determinants are evaluated by using $\zeta$-function
regularisation. It is convenient to introduce the quantity
\beq
\ca A(s)=\ze(s|H)+\ze(s|H_V)-\ze(s|A)
\:,\label{A}\eeq
where $\ze(z|L)$ is the zeta-function associated with the elliptic
operator $L$.
Thus
\beq
a(H,H_V)=\ca A'(0)=\lim_{s\to0}
\frac{d}{ds}\left[\ze(s|H)+\ze(s|H_V)-\ze(s|A)\right]\:.
\eeq

Now we also suppose $V$ to be a small perturbation potential.
We indicate by $G=H^{-1}$ the inverse operator of $H$ and by $G_V$ the
inverse of $H_V$. One has the well known operatorial equation
$G_V=G-GVG_V$, whose solution is formally given by
\beq
G_V=G\:\sum_{n=0}^\ii (-VG)^n\:.
\eeq

To begin with, let us consider the special, but physically important case
of a constant $V$. Then $H$ and $H_V$ are commuting operators, and
we may compute the complex power by means of the binomial expansion,
\beq
G_V^s=G^s\aq 1+\sum_{n=1}^\ii\:f_n(s)V^nG^n\cq\:,
\hs f_n(s)=\frac{(-1)^n\Ga(s+n)}{n!\Ga(s)}\:.
\eeq
Using the definition of the zeta-function and the properties
of the trace, we obtain
\beq
\ze(s|H_V)\equiv\Tr G_V^s=\ze(s|H)
+\sum_{n=1}^\ii\:f_n(s)\Tr(V^nH^{-s-n})
\eeq
and, in a similar way,
\beq
\ze(s|A)=\ze(2s|H)
+\sum_{n=1}^\ii\:f_n(s)\Tr(V^n H^{-2s-n})\:.
\eeq
Finally
\beq
\ca A(s)=\sum_{n=1}^\ii\:f_n(s)\aq\ze(s+n|H)-\ze(2s+n|H)\cq\:
\tr V^n\:,
\label{AForm}
\eeq
where $\tr$ is the trace on internal indices. Under the assumptions above,
all operators here involved are invertible. If $H$ has a zero
eigenvalue this must be excluded from the definition of the
zeta functions for the operators $H$ and $A$, but it gives a contribution
to $\ze(s|H_V)$. Such a contribution modifies the anomaly by an
additive logarithmic term since, in this case,
\beq
\ca A(s)=g_0 V^{-s}+
\sum_{n=1}^\ii\:f_n(s)\aq\ze(s+n|H)-\ze(2s+n|H)\cq\:
\tr V^n\:,
\label{AForm0}
\eeq
$g_0$ being the number of zero modes.

As is well known \cite{seel67-10-172}, for elliptic operators
the zeta function has only simple poles. Then the multiplicative anomaly
assumes the form
\beq
a(H,H_V)=\ca A'(0)=-g_0\ln V+
\sum_{n\geq2}\:(-1)^n\:\frac{\ga+\psi(n)}{2n}
\:\ap\Res\ze(s|H)\right|_{s=n}
\tr V^n\:,
\label{AnFF}
\eeq
where
\beq
\ga+\psi(n)=\sum_{k=1}^{n-1}\frac1k\:,
\eeq
the sum in Eq.~(\ref{AnFF}) being extended to all positive integers
where the zeta-function has a simple pole.
Here $\ga$ is the Euler-Mascheroni constant,
while $\psi(x)$ is the digamma function.
This expression has limited validity, mainly because it assumes that
 the operators involved in the product commute, but nevertheless, one should
note that it holds
as soon as one can define the zeta-functions involved. That includes,
 for example, the case
when the manifold is non-compact or when it is compact and
with boundary.

Furthermore, when one deals with the last case, i.e. a compact
$D$-dimensional manifold
with boundary,  one can certainly proceed, since the meromorphic structure
 of the
zeta function corresponding to an elliptic operator $H$ of order $h$ is well
  known.
In fact,  the short-$t$ heat-kernel asymptotic expansion of $\Tr e^{-tH}$ is
known
(see, for example, \cite{grei71-41-163}). Making then use of the Mellin
transform
\beq
\ze(s|H)=\frac{1}{\Ga(s)}\int_0^\ii dt t^{s-1}\Tr e^{-tH}
\:,\eeq
a standard procedure yields the analytical continuation
\beq
\ze(s|H)=\frac1{\Ga(s)}\sum_n\frac{K_n}{s+\frac{n-D}h}
+\mbox{ analytic part}\:,
\label{MSZF}
\eeq
where
\beq
K_n=\frac1{(4\pi)^{d/h}} \int k_n(x|H)d^Dx\:
\eeq
are the heat-kernel coefficients related to $H$, which enter in the
asymptotic expansion of $\Tr e^{-tH}$. They are, in principle,
computable (see, for example, for the case of a Laplace-type operator,
 the recent paper \cite{kirs98-15-5} and
references therein). From this we immediately obtain
\beq
\ap\Res\ze(s|H)\right|_{s=n}=\frac{K_{D-hn}(H)}{(n-1)!}
\eeq
and thus,
\beq
a(H,H_V)=-g_0\ln V+
\sum_{n\geq 2,hn\leq D}\:(-1)^n\frac{\ga+\psi(n)}{2n!}\:
K_{D-hn}(H)\:\tr V^n\:.
\label{AnFF1}
\eeq
We close this section with some remarks. The multiplicative anomaly
formula above is valid only for constant $V$. For a manifold without
boundary one has $K_r=0$, when $r$ is odd. Thus,
for a second order differential operator ($h=2$)
in any odd-dimensional  compact manifold without boundary, the multiplicative
anomaly  vanishes. It is also vanishing in two dimensions,
but it is actually present for $D=4$, being
proportional to the Seeley-DeWitt coefficient $K_0$.
In the case of first order differential operators ($h=1$),
the anomaly is non-vanishing for $D=2$,
and for the physically more interesting case $D=4$ one has
(assuming zero modes to be absent)
\beq
a(H,H_V)=\frac{1}{4}K_2(H)\tr V^2+\frac{11}{288}K_0(H)\tr V^4
\:.\label{dd}\eeq
We note the presence of the first non trivial
Seeley-DeWitt coefficient $K_2(H)$, which for a Dirac-like operator
depends on the scalar curvature of the manifold one is dealing with
and $V$ may be a mass difference. This could be interpreted
as a (potentially) interesting, finite effect of induced gravity
--according to Sakharov-- caused by the quantum
fluctuation of the matter spinor field.

It is also interesting to note that the linear term in $V$ does never
contribute to the multiplicative anomaly.
If one invokes dimensional arguments, this fact is not really surprising
in the case of constant $V$. However, as we shall see in the following,
this result is also true for an arbitrary potential. To show this important
fact we shall make use of standard perturbation theory.

\ss{Explicit expression of the anomaly from perturbation theory}
\label{S:EEAPT}

As above, we assume $H$ to be an elliptic differential operator
of known, non-degenerate spectral decomposition
$\left\{ \lambda_i, \varphi_i \right\}_{i\in I}$
and $V$ a small (non constant in general)
 perturbation  potential. Let us denote by $\mu_i$
and $a_i$ the eigenvalues of the operator $H_V$ and $A$ respectively.
By using perturbation theory up to second order in $V$, we have
\beq
\mu_i&=&\la_i+V_{ii}+\sum_{j\neq i}
\frac{V_{ij}V_{ji}}{\la_i-\la_j}+O(V^3)\:,
\\
a_i&=&\la_i^2+\la_iV_{ii}+\sum_{j\neq i}
\frac{\la_i\la_jV_{ij}V_{ji}}{\la_i^2-\la_j^2}+O(V^3)\:,
\\
V_{ij}&=&\at\ph_i,V\ph_j\ct\:.
\eeq
Let us assume that $\Re s$ is sufficiently large.
Then,  by definition $\ze(s|H)=\sum_i\lambda_i^{-s}$ while,
using again the binomial expansion
\beq
\ze(s|H_V)&=&\sum_i\mu_i^{-s}
=g_0V^{-s}+\ze(s|H)-s\sum_i\:V_{ii}\lambda_i^{-s-1}\\&&\hs
+\frac{s(s+1)}{2}\sum_i\:V_{ii}^2\lambda_i^{-s-2}
-s\sum_{i,j,i\neq j}\frac{V_{ij}V_{ji}}{\la_i-\la_j}
\lambda_i^{-s-1}+O(V^2)\:,
\eeq
\beq
\ze(s|A)&=&\sum_i a_i^{-s}
=\ze(2s|H)-s\sum_i\:V_{ii}\lambda_i^{-2s-1}\\&&\hs
+\frac{s(s+1)}{2}\sum_i\:V_{ii}^2\lambda_i^{-2s-2}
-s\sum_{i,j,i\neq j}\frac{\la_jV_{ij}V_{ji}}{\la_i^2-\la_j^2}
\lambda_i^{-2s-1}+O(V^2)\:.
\eeq
As a result, the quantity $ \ca A(s)$ can be written as
\beq
\ca A(s)&=&g_0V^{-s}+2\ze(s|H)-\ze(2s|H)+s[\Phi_1(2s)-\Phi_1(s)]
\nn\\
&&\hs -\frac{s(s+1)}{2}[\Phi_2(2s)-\Phi_2(s)]
+s[\Psi_1(2s)-\Psi_1(s)+\Psi_2(2s)]\:,
\label{Ian}
\eeq
where we have introduced the functions
\beq
\Phi_k(s)&=&\sum_i V_{ii}^k\la_i^{-s-k}
\:,\\
\Psi_1(s)&=&\sum_{i,j,i\neq j}
\frac{V_{ij}V_{ji}}{\la_i-\la_j}\la_i^{-s-1}
\:,\\
\Psi_2(s)&=&\sum_{i,j,i\neq j}
\frac{V_{ij}V_{ji}}{\la_i^2-\la_j^2}\la_i^{-s} \:.
\label{psi2}
\eeq
Since we are interested in the derivative at $s=0$,
let us assume that the behaviour of the analytic continuation of these
 functions, in a neighborhood of the origin, has the form
\beq
\Phi_k(s)&=&\frac{B_k}{s}+A_k+O(s)
\:,\\
\Psi_k(s)&=&\frac{C_k}{s}+D_k+O(s), \qquad k=1,2 \:.
\label{psi3}
\eeq
in agreement with the fact that the zeta function has only simple poles. From
Eqs.~(\ref{Ian})-(\ref{psi3}), we easily obtain
\beq
a(H,H_V)=\frac{B_2}4+D_2,
\eeq
again up to second order of perturbation theory.
As anticipated before, at first order in $V$ there
is no contribution to the multiplicative anomaly (observe that
the potential here is {\it not} necessarily constant).

When  $V$ is  constant, one has
\beq
\Phi_k(s)=V^k\ze(s+k|H)\:,
\hs\hs\Psi_1(s)=\Psi_2(s)=0\:
\eeq
and then
\beq
a(H,H_V)=\frac{V^2}{4(4\pi)^{D/h}}\:K_{D-2h}(H)\:,
\label{anom-const}
\eeq
which is the leading term in $V$ in Eq.~(\ref{AnFF}).
To summarize, up to second order of perturbation theory and for the
special class of product of two operators,
the multiplicative anomaly depends only on $V^2$
and not on the derivatives of $V$.
Moreover, for dimensional reasons, in a 4-dimensional manifold and for a
second order differential operator, the second
order approximation gives the exact result and thus,
in such  case one may argue that the form of the anomaly reads
\beq
a(H,H_V)=\frac1{4(4\pi)^{2}}
\int k_{0}(x|H)\:\tr(V^2)\:d^4x\:,
\label{anom-loc1}
\eeq
whatever be the potential (e.g. not necessarily constant).
We will confirm this result making use of a more powerful technique we
are going to introduce in the next section.

\s{The Wodzicki residue}
\label{S:WR}

For the reader's convenience, we will update in this section some basic
information concerning the Wodzicki residue  \cite{wodz87b} (see
also \cite{kass89-177-199} and the references to Wodzicki
 quoted therein)
that will be  used in the rest of the paper.

Let us consider a D-dimensional, smooth (compact) manifold without boundary, $M_D$,
and a (classical) $\Psi$DO, $Q$,
 acting on sections of vector
bundles on $M_D$. To any classical $\Psi$DO, it corresponds a complete
symbol $\si(Q)=Q(x,k)=e^{ikx}Qe^{-ikx}$, such that, modulo infinitely
smoothing operators, one has
\beq
(Q f)(x)\sim \int_{R^D}\frac{dk}{(2\pi)^{D}}\int_{R^D}dy \
e^{i(x-y)k}Q(x,k)f(y)
\:.\label{sy}\eeq
The complete symbol of $Q$ admits an asymptotic expansion for $|k| \to \ii$,
given by
\beq
Q(x,k)\sim \sum_{j=0}^\infty Q_{q-j}(x,k)
\:,\label{sy1}\eeq
where the coefficients fulfill the homogeneity property
$Q_{q-j}(x,tk)= t^{q-j}Q_{q-j}(x,k)$,
for $t>0$, being $Q_{q}(x,k) \neq 0$. The number $q$ is
called the order of $Q$. We shall deal mainly
 with self-adjoint operators and thus we  will
assume that their complex powers, beside the semigroup property, also
satisfy $(A^c)^{-s}=A^{-cs}$, the $c$ and $s$ being arbitrary complex numbers.
As an example that we will encounter, let us consider $\ln A$,
where $A$ is an elliptic operator of order $a$. Then
\beq
\si(\ln A)\sim a \ln |k|+\sum_{j=0}^\infty A_{-j}(x,k)
\:.\label{siq}\eeq
Observe that  $\ln A$ is not a classical $\Psi$DO operator, since in
general its symbol differs from that of a zero order
operator in the presence of the $\ln |k|$ term.

In order to introduce the definition of the non-commutative residue of $Q$, let
us consider an elliptic operator $A$, with $a>q$, and form the family of
$\Psi$DOs
$A_Q(u)=A+uQ$, $u$ being a real parameter. The associated zeta-function
reads
\beq
\ze(s|A_Q(u))=\Tr (A+uQ)^{-s}
\:.\label{zq}\eeq
The meromorphic structure of the above zeta-function can be obtained
from the short-$t$ asymptotics of $\Tr e^{-A_Q(u)}$ \cite{duis75-29-39}, namely
\beq
\Tr e^{-tA_Q(u)}\simeq \sum_{j=1}^\ii \al_j(u)
t^{(j-D)/a}+\sum_{k=1}^\ii \be_k(u)t^k \ln t
\:.\label{tas}\eeq
Note the presence of logarithmic terms  that lead, 
using the Mellin transform, to  double poles
in the meromorphic expansion of $\ze(s|A_Q(u))$, i.e.
\beq
\ze(s|A_Q(u))&=&\frac{1}{\Ga(s)}\at \int_0^1 +\int_1^\ii \ct dt \ t^{s-1}\Tr
e^{-A_Q(u)}\nn \\
&=&\frac{1}{\Ga(s)}\aq
\sum_{j=1}^\ii\frac{\al(u)_j}{s+\frac{j-D}{a}}-\sum_{k=1}^\ii
\frac{\be_k(u)}{(s+k)^2}+J(s,u)\cq
\:,\label{mero}\eeq
where $J(s,u)$ is the analytic part. Taking the derivative with
respect to $u$ and then the limit $u \to 0$, one gets
\beq
\lim_{u \to 0}\frac{d}{du} \Tr (A+uQ)^{-s}&=&-s\Tr (QA^{-s-1})\nn \\
&=&\frac{1}{\Ga(s)}\aq
\sum_{j=1}^\ii\frac{\al'_j(0)}{s+\frac{j-D}{a}}-\sum_{k=1}^\ii
\frac{\be'_k(0)}{(s+k)^2}+J'(s,0)\cq
\:.\label{www}\eeq
By definition, the non-commutative residue of $Q$ is
\beq
\mbox{res}(Q)=\Res \aq a \lim_{u \to 0}\frac{d}{du} \Tr (A+uQ)^{-s}  \cq_{s=-1}
=a\be'_1(0)
\:,\label{ncr}\eeq
where $\Res$ is the usual Cauchy residue. It is possible to show
that $\mbox{res}(Q)$ is independent on the elliptic operator $A$ and
that it is a trace in the algebra of classical $\Psi$DOs (actually, the
only trace up to multiplicative constants). From the
above definition and taking the derivative with respect to $u$ at $u=0$
of Eq.~(\ref{tas}),
one obtains a possible way to compute the non-commutative residue.
In fact
\beq
\Tr \at Q e^{-tA} \ct \simeq -\sum_{j=1}^\ii \al'_j(0)
t^{(j-D)/a-1}-\frac{\mbox{res}(Q)}{a} \ln t  +O(t\ln t)
\label{bubu}\eeq
and so the non-commutative residue of $Q$ can be read off from the
short $t$ asymptotics of the quantity $\Tr\at Qe^{-tA}\ct$,
just picking the coefficient associated with $\ln t$.
When the manifold is non-compact, this is
one of the methods that we have at hand for evaluating the Wodzicki residue,
as long as all the traces involved exist.

For the case of a compact manifold, Wodzicki has obtained a useful
local form of the non-commutative residue, that is,
a density  which can be integrated to yield the non-commutative residue,
namely
\beq
\mbox{res}(Q)=\int_{M_D}\frac{dx}{(2\pi)^{D}}\int_{|k|=1}Q_{-D}(x,k)dk
\:.\label{wod2}\eeq
Here the component of order $-D$ (remember that $D$ is the dimension of
the manifold) of the complete symbol appears.

Let us now consider an elliptic (self-adjoint) operator $B$ of order
$b>q$. From
Eq.~(\ref{www}) and the requirement $\Tr [Q (B^c)^{-z}]=\Tr (Q
B^{-cz})$.  One has the following

\bigskip\noindent{\bf Lemma 1.} In a neighborhood of $z=0$,
\beq
\Tr (Q B^{-z})= \frac{\mbox{res} (Q)}{z b}
+\frac{\ga\:\mbox{res} (Q)}{b}-R_Q(B)+O(z)
\:,\label{A1}\eeq
where $\ga$ is the Euler-Mascheroni constant and the quantity $R_Q(B)$
satisfies the relation
\beq
R_Q(B^c)-R_Q(B)=\frac{\ga\:\mbox{res} (Q)}{b}\frac{1-c}{c}\:,
\hs c>0\:.\label{cog}
\eeq
The latter equation gives another way to compute the non-commutative residue,
namely
\beq
\mbox{res} (Q)=b \aq \Res \Tr (Q B^{-z}) \cq_{z=0}
\:.\label{wodf}\eeq

Making use of Lemma 1, one also obtains (with $c_1 >0$ and $c_2 >0$):
\beq
\Tr \left[Q (A^{-c_1z}- B^{-c_2z})\right]&=& \frac{\mbox{res}
(Q)}{z}\at\frac{1}{ac_1}-\frac{1}{bc_2} \ct
+\ga\:\mbox{res} (Q)
\at\frac{1}{a}-\frac{1}{b}\ct \nn \\
&&\hs\hs-R_Q(A)+R_Q(B)+O(z)
\:.\label{ccc}\eeq
In particular, if $ac_1=bc_2=1$ there is no pole at $z=0$, and then
\beq
\Tr \left[Q (A^{-\fr{z}{a}}- B^{-\fr{z}{b}})\right]=
\ga\:\mbox{res} (Q)\at\frac{1}{a}-\frac{1}{b}\ct
-R_Q(A)+R_Q(B)+O(z)\,.
\label{mmm}
\eeq
On the other hand, we also have  \cite{kont95b}

\bigskip\noindent{\bf Lemma 2}
\beq
\Tr \left[Q (A^{-\fr{z}{a}}- B^{-\fr{z}{b}})\right]
=-\mbox{res} \aq Q \at \frac{\ln
A}{a}-\frac{\ln B}{b}\ct \cq+O(z)
\,.
\label{mmm1}
\eeq
This Lemma is a consequence of Lemma 1. In fact, if $C$ is a
suitable elliptic operator, making use of the relation
$C^z=I+z\ln C+O(z^2)$, one obtains
\beq
\Tr \left[Q (A^{-\fr{z}{a}}- B^{-\fr{z}{b}})\right]&=&\Tr \left[
Q (A^{-\fr{z}{a}}-
B^{-\fr{z}{b}})C^z C^{-z}\right] \nn \\
&=&-z\Tr \left[Q \left(\frac{\ln A}{a}-
\frac{\ln B}{a}\right) C^{-z}\right]+O(z^2)
\:.\label{bbb}\eeq
Now $Q (\frac{\ln A}{a}-\frac{\ln B}{a})$ is a classical $\Psi$DO  and
we can make use of Lemma 1 to obtain the desired result.
It is easy to show that Lemmas 1 and 2 allow us to rewrite
Eq.~(\ref{mmm}) as
\beq
\Tr \aq Q(A^{-z}-B^{-z})\cq=
\frac{\mbox{res}(Q)}{z}
\at\frac{1}{a}-\frac{1}{b}\ct
-\mbox{res}\aq Q\at\frac{\ln A}{a}-\frac{\ln B}{b}\ct\cq+O(z)
\:.\label{l3}\eeq
As a result, we have

\bigskip\noindent{\bf Lemma 3}
\beq
\lim_{z\to0}\frac{d}{dz}\ag z
\Tr\aq Q(A^{-z}-B^{-z})\cq\cg=
-\mbox{res} \aq Q\at\frac{\ln A}{a}-\frac{\ln B}{b}\ct\cq
\:.\label{ll3}\eeq

\s{The multiplicative anomaly formula}
\label{S:MAF}

In this section we will present a quick proof of  the multiplicative
anomaly formula, following the derivation in \cite{kont95b}.
We consider two invertible, elliptic, self-adjoint
operators $A$ and $B$ on $M_D$ of orders $a$ and $b$
respectively, and the quantity
\beq
F(A,B)=\frac{\det (AB)}{(\det A)( \det B)}=e^{a(A,B)}
\:.\label{a4}\eeq
Moreover, we introduce the family of  $\Psi$DOs
\beq
A(t)&=&\eta^t B^{\fr{a}{b}}\,,\hs A(0)=B^{\fr{a}{b}}\:,
\hs A(1)=A\:,\nn\\
\eta&=&AB^{-\fr{a}{b}}\,,\hs\mbox{ord}\:\eta=0,
\hs\mbox{ord}\:A(t)=a,\hs\mbox{ord}\:(A(t)B)=a+b\:.
\eeq
and the function
\beq
F(A(t),B)=\frac{\det (A(t)B)}{(\det A(t))( \det B)}
\:.\label{a5}\eeq
One trivially gets
\beq
F(A(0),B)=1
\,, \hs F(A(1)),B)=\frac{\det (AB)}{(\det A)(\det B)}=F(A,B)
\:.\label{a6}\eeq
As a consequence, one  is led to deal with  the following expression
for the anomaly
\beq
a(A(t),B)=\ln F(A(t),B)=-\lim_{s \to 0} \ \partial_s \aq \Tr (
A(t)B)^{-s}-\Tr A(t)^{-s}-\Tr B^{-s} \cq
\:.\label{a7}\eeq
This quantity has the properties: $a(A(0),B)=0$ and $a(A(1),B)=
a(A,B)$.

The next step is to
compute the first derivative of $a(A(t),B)$ with respect to $t$,
the result being
\beq
\partial_t a(A(t),B)=\lim_{s\to0}
\partial_s\ag s\Tr\ln\eta\aq(A(t)B)^{-s}
-\Tr A(t)^{-s}\cq\cg
\:.\label{a8}\eeq
Furthermore, making  use of the Lemma 3, i.e. Eq.~(\ref{ll3}), with $Q=\ln \eta$, one  obtains
\beq
\partial_t a(A(t),B)= \mbox{res} \aq \ln \eta \at  \frac{\ln A(t)}{a}-
\frac{\ln [A(t)B]}{a+b}\ct \cq
\:.\label{a9}\eeq
Finally, performing the integration with respect to $t$
from $0$ to $1$, one
gets  the Kontsevich-Vishik multiplicative anomaly formula \cite{kont95b},
namely
\beq
a(A,B)=\int_0^1 dt \  \mbox{res} \aq \ln \eta \at \frac{\ln A(t)}{a}-
\frac{\ln [A(t)B]}{a+b}\ct \cq
\:.\label{a10}\eeq

The multiplicative anomaly formula, Eq.~(\ref{a10}), notably simplifies
in the special case of commuting operators. In fact one has
\beq
\frac{\ln [A(t)B]}{a+b}=\frac{t\ln\eta}{a+b}
+\frac{\ln B}{b}\,,\hs
\frac{\ln A(t)}{a}=\frac{t\ln\eta}{a}+\frac{\ln B}{b}
\label{a11}\eeq
and
\beq
\frac{\ln [A(t)B]}{a+b}-\frac{\ln A(t)}{a}
=t\frac{bt\ln\eta}{a(a+b)}
\:.\label{a111}\eeq
As a result,
\beq
a(A,B)=
\frac{b}{2a(a+b)}\mbox{res}\aq (\ln(AB^{-\fr{a}{b}}))^2
\cq
\:,\label{wod3}\eeq
which can be rewritten as the Wodzicki multiplicative formula
\cite{kass89-177-199}
\beq
a(A,B)= \frac{\mbox{res}\aq (\ln(A^bB^{-a}))^2
\cq}{2ab(a+b)}=a(B,A)
\:,\label{wod33}\eeq
where the symmetry property in $A$ and $B$ is manifest.

We would like to end this section with the following simple (but
important) remark. The
notion of non-commutative residue can be introduced only for classical
$\Psi$DOs, namely the ones whose symbol admits the asymptotics
Eq.~(\ref{sy1}). For example, in general $\ln A$ with $A$ elliptic and
$a>0$, is not a classical
$\Psi$DO, but $\frac{\ln A}{a}-\frac{\ln B}{b}$ is,
as well as $\ln\eta$, since ord$(\eta)=0$. Thus, all the formulae in which the
non-commutative residue appears are well defined.

For more than two operators, the recurrence equation
\beq
a(A,B,C)=a(AB,C)+a(A,B)
\label{3a}
\eeq
holds and thus, the knowledge of the multiplicative formula
for the product of two operators is sufficient for computing
the multiplicative anomaly associated with the product of
an arbitrary number of operators.

\ss{The multiplicative anomaly formula in lower dimensions}
\label{S:MALD}

In the commutative case, examples of the evaluation of the
multiplicative anomaly have been already presented (see
\cite{eliz97u-394,eliz97u-404}). For the non commutative one, we have
obtained some insights by using perturbative arguments. Here we would
like to compute the multiplicative anomaly in the lower dimensional
cases $D\leq4$, by making use of the
Kontsevich-Vishik formula Eq.~(\ref{a10}). We shall consider a compact
manifold, in order to make use of the Wodzicki local formula for the
evaluation of the non-commutative residue. With this aim, we start from
the following
property for the product of the two non-commuting self-adjoint elliptic
$\Psi$DOs that we shall consider,  $Q$
and $P$  (Baker-Campbell-Haussdorf formula)
\beq
\ln \at PQ\ct=\ln P+\ln Q+[\ln P,G]+\frac{1}{12}[\ln Q,[\ln Q,\ln
P]]+O(\ln^3 Q\ln P)
\:,\label{bch}\eeq
where $G$ is a suitable nested commutator. Making use of this operator
identity, we have that the zero order $\Psi$DO $Q_t$, defined by
\beq
Q_t=\frac{\ln (\eta^t B^{\fr{a}{b}})}{a}
-\frac{\ln (\eta^t B^{\fr{a+b}{b}})}{a+b}
\:,\label{pt}\eeq
may be rewritten as
\beq
Q_t=-\frac{b t \ln \eta}{a(a+b)}+[\ln \eta,G]
-\frac{t}{12b}[\ln B,[\ln B,\ln\eta]]+O(\ln^3 B \ln \eta)
\:.\label{b1}\eeq
Since $\res(\ln\eta Q_t)$ is the integrand in  the multiplicative anomaly
formula,  Eq.~(\ref{a10}), and the non commutative residue is a trace
in the algebra of classical $\Psi$DOs, one has
\beq
\res\at\ln\eta[\ln\eta,G]\ct=0
\:.\label{op}\eeq
As a consequence, the Kontsevich-Vishik formula Eq.~(\ref{a10}) yields
\beq
a(A,B)&=& \frac{\mbox{res}\at \ln^2 \eta \ct}{2ab(a+b)}
-\frac{1}{12 b} \mbox{res}\at \ln \eta [\ln B,[\ln B,\ln\eta]] \ct \nn\\
&&\hs\hs+{\cal O}\at\mbox{res}\,(\ln^2\eta\ln^3 B)\ct \:,\label{wkv}
\eeq
in which the term which is non vanishing in the commutative case has been conveniently isolated.

Now we recall that for a zero order $\Psi$DO,  $\eta$, one has
$\si(F(\eta))=F(\si(\eta))$ (for $F(.)$ analytic) and,
moreover, that the symbol $\si$ of the commutator of two $\Psi$DOs
reads
\beq
\si([Q,P])&=&2\aq\frac{\partial}{\partial x^\mu}\si(Q)
\frac{i\partial}{\partial k_\mu}\si(P)
-\frac{i\partial}{\partial k_\mu}\si(Q)
\frac{\partial}{\partial x^\mu}\si(P)\cq \nn \\
&&\hs\hs+\mbox{higher order derivatives}\:.
\eeq
If we assume, as in Sect.~\ref{S:PDMA}, that $A$ and $B$ have the
same principal and sub-leading symbols except for the homogeneous one
of degree zero, namely $a=b$ and $\si(A)-\si(B)=V(x)$, then
\beq
&&\si(\ln\eta)=O(1/|k|^a)\:,\hs
\frac{\partial}{\partial x^\mu}\si(\ln\eta)=O(1/|k|^a)\:,\\
&&\frac{\partial}{\partial x^\mu}\si(\ln B)=O(|k|^0)\:,
\hs\frac{\partial}{\partial k^\mu}\si(\ln B)=O(1/|k|),
\eeq
and it follows that
\beq
\si\at\ln\eta[\ln B,[\ln B,\ln\eta]]\ct=O(1/|k|^{2a+2})
\:.\label{gu}\eeq
This means that if the dimension of the compact manifold $D$ satisfies
 $D < 2a+2$, only the first term on the right-hand side
of Eq.~(\ref{wkv}) gives a non vanishing contribution to the anomaly.
Such a condition is satisfied in the particular but important case
of second order differential operators in four dimensions.
On the other hand, for first order differential operators $a=b=1$ and  $D=4$, 
also the second term of Eq.~(\ref{wkv})
gives a contribution. This exact analysis confirms the perturbative
results of  Sect.~\ref{S:PDMA}.

\s{Heat kernel coefficients from the multiplicative anomaly}
\label{S:HKFA}

As we we have shown in Section \ref{S:PDMA}, Eq.~(\ref{AnFF1}),
in the case of constant $V$ the multiplicative anomaly $a(H,H_V)$
can be directly related to the heat kernel coefficients $K_n(H)$,
but it can also be computed by means of Wodzicki formula,
Eq.~(\ref{wod33}). 
This means that one could  use the Wodzicki residue in order to
compute heat-kernel coefficients (see Ref.~\cite{kass89-177-199,eliz97-30-2735}).
Now we illustrate the method for an invertible  
differential operator $H$,  of order $h$, on a manifold without boundary. 

For this case, Eq.~(\ref{wod33}) reduces to
\beq 
a_D(H_V,H)=\frac1{4h}\:\res\aq\ln(H_VH^{-1})\cq^2=
\frac1{4h}\:\res\aq\ln(I+VH^{-1})\cq^2\:,
\eeq
If $V$ is a constant, from Eq.~(\ref{AnFF1}) one also obtains
\beq 
a_D(H_V,H)=\frac1{4}K_{D-2h}(H)\:V^2+O(V^3)\:.
\eeq
Now, taking the limit $V\to0$, one may show that 
\beq 
K_{D-2h}(H)=\lim_{V\to0}\frac{a_D(H_V,H)}{4V^2}
=\frac1{4h}\:\res H^{-2}\:,
\eeq
in agreement with Ref.~\cite{eliz97-30-2735}.

\s{Some physical examples}
\label{S:examples}

Here we are going to consider some basic examples in which the
multiplicative anomaly may  actually have physical consequences.
Our aim has been, in fact,
to look for the most simple situations (from the point of view of
the number of space dimensions and of the order and nature of the differential
operators involved) that might already exhibit a non-zero anomaly.
It should not come as a surprise that, among them, the harmonic
oscillator
plays an important role. It yields non-trivial results easy to
calculate analytically in any number of dimensions.

\ss{Presence of the anomaly for Dirac-like operators in one space dimension}

Consider the square root of the harmonic oscillator obtained by
Delbourgo in Ref.~\cite{delb95u-56}.
This example  has potentially some interesting physical
applications, for it is well known that a fermion in an external
constant electromagnetic field has a similar spectrum (Landau spectrum).
Exactly in the same way as when
going from the Klein-Gordon to the Dirac equation and at the same price of
doubling the number of components
(e.g., introducing spin), Delbourgo has constructed a model for which
there exists a square root of its Hamiltonian, which is very close to the
one for the harmonic oscillator. It is in fact different from the
Dirac oscillator
introduced by several other authors, corresponding to the minimal
substitution $\vec{p} \rightarrow \vec{p} -i\alpha \vec{r}$. The main
difference lies in the introduction now of the parity operator, $Q$. Whereas
creation and destruction operators for the harmonic oscillator,
$a^\pm = P\pm i X$, are non-hermitian,
the combinations $\ca D^\pm=P\pm iQX$ are hermitian and
\begin{eqnarray}
H^\pm \equiv (\ca D^\pm)^2 =P^2 +X^2 \mp Q =2H_{\mbox{osc}} \mp Q.
\end{eqnarray}
Notice that the parity term commutes with $H_{\mbox{osc}}$. Doubling the
components  ($\sigma_i$ are the Pauli matrices)
\begin{eqnarray}
P\rightarrow -i\sigma_1 \frac{\partial}{\partial x}, \qquad
X\rightarrow \sigma_1 x, \qquad
Q\rightarrow \sigma_2 ,
\end{eqnarray}
the operators $\ca D^\pm$ are represented by
\begin{eqnarray}
\ca D^\pm \rightarrow -i\sigma_1
\frac{\partial}{\partial x}\pm\sigma_3 x.
\end{eqnarray}

In the sequel,  we will only consider the operator
$\ca D\equiv\ca D^+$.
It has for eigenfunctions and eigenvalues, respectively,
\beq
\psi^\pm_n (x)=\frac{-ie^{-x^2/2}}{\sqrt{2^{n+1}\,(n-1)!\sqrt{\pi}}}
\left(\begin{array}{c}-i\left[H_{n-1}(x)\pm H_n(x)/\sqrt{2n}\right]\\
\left[ H_{n-1} (x) \mp H_n(x)/\sqrt{2n}\right]\end{array}\right),
\qquad\lambda_n=\pm\sqrt{2n},\:\:n\geq1,\nonumber
\eeq
\beq
\psi_0 (x)=\frac{e^{-x^2/2}}{\sqrt{2\sqrt\pi}}
\left(\begin{array}{c} 1 \\ i \end{array} \right),
\qquad \lambda_0 =0,
\eeq
where the $H_n(x)$ are Hermite polynomials.

The two operators we shall consider for the calculation of the anomaly are
$\ca D$ and $\ca D_V=\ca D+V$, $V$ being a real,
constant potential with  $|V|<\sqrt{2}$,
that goes multiplied with the identity matrix in the two (spinorial)
dimensions (omitted here).
Notice that $\ca D$ and $\ca D_V$ are hermitian, commuting operators.
The zeta function for the operator $\ca D$ reads
\begin{eqnarray}
\zeta(s|\ca D)=\sum_i\lambda_i^{-s}
=\sum_{n=1}^\infty[1+(-1)^{-s}]
\left( \sqrt{2n}
\right)^{-s}=[1+(-1)^{-s}] 2^{-s/2}\zeta_R(s/2),
\end{eqnarray}
$\ze_R(s)$ being the usual Riemann zeta function, which has a simple pole
at $s=1$. Furthermore, the manifold is not compact, thus,
by  direct use of Eq.~(\ref{AnFF}) we readily get
\begin{eqnarray}
a(\ca D,\ca D_V)=\frac{V^2}{2}-\ln V.
\end{eqnarray}
The logarithmic term is due to the presence of a zero mode.

How does this match with the Wodzicki formula?
First of all, we point out that one has to deal with zero modes.
Thus some care must be taken to
properly treat them. Secondly, since we are working in a non-compact
manifold, we cannot used the local formula to evaluated the
non-commutative residue. A direct check, modulo the zero mode problem,
that makes use of Eq.~(\ref{bubu}), yields the multiplicative
anomaly above.
It thus seems  that Wodzicki's expression
requires only  small modification in order to deal with spinorial
operators as the ones we have here.

We conclude by pointing out that we have here, before us, the first
and most
simple example of the presence of a non-trivial anomaly for operators of
degree one in a space of dimension one (spinorial, however).

\ss{Generalization of the Dirac-like operators to $D$ dimensions}

Referring again to the work by Delbourgo in Ref.~\cite{delb95u-56},
the above operators $\ca D^+$ and $\ca D^-$ admit possible extensions
to $D$ dimensions, which have for eigenvalues, respectively,
\begin{eqnarray}
\lambda_+ =\pm\sqrt{2(2n_r+l)}, \qquad
\lambda_- =\pm\sqrt{2(2n_r+l+D)},\qquad n_r,l = 0,1,2,
\ldots  \label{6.7}
\end{eqnarray}
Notice that $\ca D^+$ exhibits a zero mode (for $n_r=l=0$), what is not
the
case with $\ca D^-$. Each of these two operators provides a different
example for the calculation of the anomaly, and we will denote the
corresponding partners by $\ca D_V^\pm$ (see the preceding subsection).

The basic zeta functions are now
\begin{eqnarray}
\zeta_+(s|H)&=& 
\left[ 1+(-1)^{-s}\right]2^{-s/2}
{\sum_{n_r,l=0}^\infty}' \left(2n_r+l\right)^{-s/2} \nn
\\&&\hs\hs=\left[ 1+(-1)^{-s}\right]2^{-s/2}
\zeta_2\left(\frac{s}{2},\left. 0\right|(2,1)\right), \\
\zeta_-(s|H)&=& 
\left[ 1+(-1)^{-s}\right]2^{-s/2}
\sum_{n_r,l=0}^\infty \left(2n_r+l+D\right)^{-s/2} \nn
\\&&\hs\hs=\left[ 1+(-1)^{-s}\right]2^{-s/2}
\zeta_2\left(\frac{s}{2},\left. D\right|(2,1)\right).
\end{eqnarray}
As usually, the prime means  in the first expression 
that the zero mode has to be excluded
from the sum (i.e., the term with $n_r=l=0$). Here
 $\zeta_2(s,b|(a_1,a_2))$ is the Barnes zeta function in 2 dimensions.
It has simple poles at the points $s=1,2$ (in general at $s=N,N-1,\ldots$
for $\zeta_N$) and the residues
are well known, as given by generalized Bernoulli
polynomials at those points (see Refs.
\cite{barn03-19-374,dowk94-35-4989,kirs96-54-4188}).
Again using Eq.~(\ref{AnFF}), one gets
\begin{eqnarray}
a_+\equiv a(\ca D^+,\ca D^+_V)&=& \sum_{j=1}^2 \frac{V^{2j}}{2^j}
\frac{\psi (2j)+\gamma}{j}\  \mbox{Res}\ \zeta_2
\left( j, \left. 0 \right| (2,1) \right)-\ln V, 
\label{a+}\\
a_-\equiv a(\ca D^-,\ca D^-_V)&=& \sum_{j=1}^2 \frac{V^{2j}}{2^j}
\frac{\psi (2j)+\gamma}{j}\  \mbox{Res}\ \zeta_2
\left( j, \left. D \right| (2,1) \right),
\label{a-}\end{eqnarray}
where, as before, the logarithmic term is due to the presence of the
zero mode, and  $\psi$ is the digamma function.
 For $D=2$, the result is
\begin{eqnarray}
 a_+ &=&\frac{V^2}{4}\left(\frac{11V^2}{12}-1\right)-\ln V, 
\nn\\
 a_- &=&\frac{V^2}{4}\left(\frac{11V^2}{12}-1\right).
\nn\end{eqnarray}
Let us now check with the result obtained from Wodzicki's formula
\begin{eqnarray}
a_W(A_1,A_2)= \frac{\mbox{res}\, \left[ \left(\ln (A_1^{a_2}A_2^{-a_1})
\right)^2\right]}{
2 a_1a_2(a_1+a_2)} = \frac{1}{4}\ \mbox{res} \left\{ \left[\ln \left(
I + \frac{V}{\ca D^\pm} \right)\right]^2\right\}.
\end{eqnarray}
Here $a_1=a_2=1$. Looking for the non-commutative residue,
making use again of Eq.~(\ref{bubu}), of course one obtains
Eqs.~(\ref{a+}) and (\ref{a-}) (modulo the
logarithmic term corresponding to the zero mode, which is
immediate to supply). From the physical point of view,
it is not completely clear if these operators make sense for any value
of $D$. In fact, in Ref.~\cite{delb95u-56}
some doubts were arisen
concerning their precise physical meaning in four and higher
dimensions.

\ss{Harmonic oscillators in $D$ dimensions}

Let us recall the case of the harmonic oscillators in $D$ dimensions,
with angular frequencies $(\omega_1,\ldots,\omega_D)$.
The eigenvalues read
\beq
\lambda_{\bar n}=\bar n\cdot\bar\om+b\:,
\hs\bar n\equiv(n_1,...,n_D)\:,
\hs\bar\om\equiv(\om_1,...,\om_D)\:,
\hs b=\frac12\sum_{k=1}^D\om_k
\eeq
and the related zeta function is the Barnes one $\ze_D(s,b|\bar\om)$,
whose poles are to be found at the points $s=k$ ($k=D,D-1,\ldots,1$).
Their corresponding residua can be expressed  in terms of
generalized Bernoulli
polynomials $B^{(D)}_{D-k}(b|\bar\om)$. They are defined by
\cite{norl22-43-21}
\beq
\frac{t^D e^{-at}}{\prod_{i=1}^D \left( 1- e^{-b_it} \right)} =
\frac{1}{\prod_{i=1}^D b_i} \sum_{n=0}^\infty
B_n^{(D)} (a|b_i) \frac{(-t)^n}{n!}.
\eeq
The residua of the Barnes zeta function are then:
\beq
\mbox{Res} \, \zeta_D (k,b|\bar\omega) = \frac{(-1)^{D+k}}{(k-1)!
(D-k)!\, \prod_{j=1}^D \omega_j} B_{D-k}^{(D)} (b|\bar\omega), \qquad
k=D,D-1, \ldots.
\eeq
Now, if $V$ is a constant potential,
from Eq.~(\ref{AnFF}) we easily obtain
\begin{eqnarray}
a(H,H_V)=\frac{(-1)^D}{2\prod_{j=1}^D \omega_j}\sum_{k=1}^{[D/2]}
\frac{[\ga+\psi(D-2k)] \, B_{2k}^{(D)}(b|\bar\omega)}
{(2k)!\,(D-2k)!}\:V^{2k}.
\end{eqnarray}

Notice that in our case the generalized Bernoulli polynomials
of odd order vanish: $B_{1}^{(D)}(b|\bar{\omega})$ $=
B_{3}^{(D)}(b|\bar{\omega})$ $= \cdots =0$,
for any $D$. On the other hand, the remaining generalized Bernoulli
polynomials are {\it never} zero, in fact
\beq
B_0^{(D)} (b|\bar\omega) &=& 1,\hs\hs
B_2^{(D)} (b|\bar\omega)=-\frac{1}{12} \sum_{i=1}^D \omega_i^2, \nn \\
B_4^{(D)} (b|\bar\omega) &=& \frac{1}{24} \left[ \frac{7}{10}  \sum_{i=1}^D
 \omega_i^4 +\sum_{i<j} \omega_i^2\omega_j^2\right],  \\
B_6^{(D)} (b|\bar\omega) &=& -\frac{5}{96} \left[ \frac{31}{70}  \sum_{i=1}^D
 \omega_i^6 + \frac{7}{10}\sum_{i\neq j} \omega_i^4\omega_j^2
+ \sum_{i<j<k} \omega_i^2\omega_j^2 \omega_k^2 \right], \quad \ldots \nn
\eeq
As a consequence, the anomaly does not  vanish in any case, not for
$D$ odd or $D=2$, whatever the frequencies $\omega_i$ be.
Moreover, only even powers of the potential $V$ appear.
Again, since the manifold is not compact, a direct use of Eq.~(\ref{bubu})
confirms the validity of the Wodzicki formula.

\s{Conclusion}
\label{S:conclusion}

In this paper, we have considered  the issue of the evaluation of
functional determinants that typically are present in the one-loop
approximation of the Euclidean  partition function. We have stressed that
the regularised determinant of a
product of elliptic operators has a multiplicative anomaly factor.
In general, such a factor is not vanishing and
it should be taken into account in associated physical applications,
since it can play an important role.
The non-commutative case of the multiplicative anomaly
formula has been discussed in some detail, making use of different
techniques. We have started with an elementary application of
standard perturbation theory of linear operators to a particular
case involving a product of two of them.

Then, more powerful techniques of
$\Psi$DOs and the Wodizcki theory of the non-commutative
residue  have been employed in a useful
re-derivation of the general multiplicative anomaly formula.

One of the most interesting results obtained in the paper deals with the
possibility
to notably simplify the general formula for the non-commuting case in
the lower dimensional situations: $D=2$, $D=4$, both of which turn out
to be very relevant for physical applications.

Several basic examples have been discussed with care,
mainly in the non compact case. Here,  another result of
our paper is the formulation of a conjecture that extends the
validity of the Wodzicki
formula to the calculation of the multiplicative anomaly in the case of
non-compact manifolds.

\ack{We would like to thank Klaus Kirsten for valuable discussions.
This work has been supported by the cooperative agreement
INFN (Italy)--DGICYT (Spain).
EE has been financed also by DGICYT (Spain), project PB96-0925,
and by  CIRIT (Generalitat de Catalunya),  grant 1995SGR-00602.
}

\end{document}